\documentclass[showkeys,showpacs,superscriptaddress]{revtex4}
\usepackage{graphicx,color}
\usepackage{amsfonts}
\usepackage{amssymb}
\usepackage{amsmath}
\usepackage{xcolor}

\begin{document}

\title{
Codimension-1 thick brane solutions in higher-dimensional $R^n$ gravity
}

\author{
V. Dzhunushaliev
}
\email{v.dzhunushaliev@gmail.com}
\affiliation{
Department of Theoretical and Nuclear Physics,  Al-Farabi Kazakh National University, Almaty 050040, Kazakhstan
}
\affiliation{
Institute of Experimental and Theoretical Physics,  Al-Farabi Kazakh National University, Almaty 050040, Kazakhstan
}
\affiliation{Academician J.~Jeenbaev Institute of Physics of the NAS of the Kyrgyz Republic, 265 a, Chui Street, Bishkek 720071, Kyrgyzstan
}

\author{V. Folomeev}
\email{vfolomeev@mail.ru}
\affiliation{
Institute of Experimental and Theoretical Physics,  Al-Farabi Kazakh National University, Almaty 050040, Kazakhstan
}
\affiliation{Academician J.~Jeenbaev Institute of Physics of the NAS of the Kyrgyz Republic, 265 a, Chui Street, Bishkek 720071, Kyrgyzstan
}
\affiliation{
	National Nanotechnology Laboratory of Open Type,  Al-Farabi Kazakh National University, Almaty 050040, Kazakhstan
}

\author{A. Serikbolova}
\affiliation{
Department of Theoretical and Nuclear Physics,  Al-Farabi Kazakh National University, Almaty 050040, Kazakhstan
}

\begin{abstract} Vacuum plane symmetric solutions within $\alpha R^n$ modified gravity are obtained. The  solutions can be regarded
as describing thick branes with codimension 1 in a higher-dimensional spacetime.
The dependence of the solutions on the values of four free parameters of the system under consideration is studied numerically.
It is shown that when the parameter $\alpha$ increases, the solutions tend to saturation.
We also show that in all cases under consideration the spacetime is asymptotically anti-de Sitter one.
\end{abstract}


\maketitle

\section{Introduction}

It is now believed that $D$-branes may play an important role in string theory~\cite{Hashimoto:2012vsa}.
Such branes are in general multidimensional surfaces to which end points of strings can be attached.
It was shown in Ref.~\cite{Polchinski:1995mt} that $D$-branes are in fact solutions
of the type black p-brane solutions in supergravity. In turn, p-branes are extended objects in low-energy string theory
possessing an event horizon~\cite{Horowitz:1991cd}. Such solutions are obtained in the presence of scalar and tensor fields.
Naturally, the question arises whether it is possible to obtain {\it vacuum} $D$-brane solutions. In the present paper,
we study this issue and show that regular vacuum brane solutions may exist in multidimensional modified theories of gravity.

Among $D$-branes, one can consider the case of a brane in the from of a spatially three-dimensional surface.
This is the so-called braneworld scenario, within which it is assumed that our Universe is such a brane embedded in some bulk space.
At the present time, such scenario is used quite widely in solving some known problems in high-energy physics
 (see, e.g., Refs.~\cite{arkani,gog,rs}). In particular, this refers to
the fermion generation puzzle \cite{fermi1,fermi2} and the nature of dark energy \cite{de} and dark matter \cite{dm}.

In Einstein's general relativity, there are many solutions describing various types of $D$-branes.
For instance, these can be the so-called thick branes~\cite{Dzhunushaliev:2009va}.
To the best of our knowledge, to obtain all such solutions, the presence of matter sources is a necessary condition.
From the physical point of view, this is because in general relativity regular solutions can almost always be obtained only
in the presence of some sources. For example, these are solutions with scalar~\cite{Schunck:2003kk} or vector and spinor~\cite{Herdeiro:2017fhv,Dzhunushaliev:2019kiy} fields.
A natural question to ask is whether it is possible to find regular vacuum brane solutions.
Within the framework of modified gravity, it was shown in Refs.~\cite{Dzhunushaliev:2009dt,Zhong:2015pta}
that in a five-dimensional spacetime there can exist a four-dimensional thick brane described by a regular vacuum solution
(see also Ref.~\cite{Dzhunushaliev:2019wvv} where thick brane solutions within higher-dimensional modified gravity have been considered).
In the present study we show that such codimension-1 thick brane solutions can exist in a space of arbitrary dimension within
the framework of modified gravity.

Modified theories of gravity are alternative models to explain the current accelerated expansion of the Universe
(for a general review on the subject, see, e.g., Refs.~\cite{Nojiri:2010wj,DeFelice:2010aj,Nojiri:2017ncd}).
In the simplest case a modification of general relativity is performed by mean of the replacement of the Einstein gravitational Lagrangian $\sim R$ by the
modified Lagrangian~$\sim~F(R)$, where $F(R)$ is some function of the scalar curvature $R$.
From the mathematical point of view, the field equations, which are obtained by varying the modified action with respect to the metric,
have a more rich structure of possible solutions; this permits one to apply them to get new physical results.

In the present paper, we show that thick $D$-branes with codimension 1 can be described by regular vacuum solutions obtained within modified gravity.
This means that to construct such $D$-branes one need not use matter sources. For this purpose, in Sec.~\ref{eqs} we write down the corresponding
equations for  $F(R)\sim R+R^n$ modified gravity. In Sec.~\ref{num_sol} these equations are solved numerically
for different values of system parameters. Finally, in Sec.~\ref{conclusion} we summarize the results obtained.

\section{Field equations in $R^n$ modified gravity}
\label{eqs}

We consider a thick brane with codimension 1 in a $N$-dimensional spacetime. The corresponding gravitational action can be written in the form
[the metric signature is $(+,-,-,\ldots)$]
\begin{equation}
\label{1_10}
   S = \int d^N x\sqrt {^{(N)}G} \left[ - \frac{R}{2} + f(R) \right]~,
\end{equation}
where $f(R)$ is an arbitrary function of the scalar curvature $R$; $G$ is the determinant of  the $N$-dimensional metric
$G_{AB}$ (hereafter capital Latin indices refer to a multidimensional spacetime: $A, B = 0, 1, \dots , N $).
Variation of the action~\eqref{1_10} with respect to the metric $G_{AB}$ yields the equations of modified gravity
\begin{equation}
\label{1_20}
   R_{A}^B - \frac{1}{2} \delta_{A}^B R = \hat{T}_{A}^B,
\end{equation}
whose right-hand side is 
\begin{equation}
\label{1_30}
	\hat{T}_{A}^B = -\left[ \left(
   	\frac{\partial f}{\partial R}
  \right) R_{A}^B - \frac{1}{2}\delta_{A}^{B} f + \left(
  		\delta_{A}^{B} g^{LM} - \delta_{A}^{L} g^{BM}
  \right)
  \left( \frac{\partial f}{\partial R}\right)_{;L;M}
  \right],
\end{equation}
where the semicolon denotes the covariant derivative.

In the present paper, we consider the following special choice of modified gravity:
\begin{equation}
\label{1_40}
   f(R)=-\alpha R^n,
\end{equation}
where $\alpha>0$ and $n$ are constants. Since here we seek codimension-1 thick brane solutions in a $N$-dimensional spacetime, the metric can be taken in the form
\begin{equation}
	ds^2 = e^{2\beta(x^N)} \left[
		(dx^0)^2 - (dx^1)^2 - \dots -(dx^{N - 1})^2
	\right] - (dx^N)^2 ,	
\label{1_50}
\end{equation}
where the metric function $\beta$ depends only on one coordinate $x^N$.
The corresponding components of the Ricci tensor are
\begin{eqnarray}
\label{1_60}
	R_{00} &=& e^{2\beta} \beta^{\prime\prime} + N {\beta^{\prime}}^2
	\epsilon^{2\beta},
\\
	R_{CC} &=& - e^{2 \beta} \left(
		\beta^{\prime\prime} + N {\beta^{\prime}}^2
	\right), \quad \text{where} \quad C = 1, 2, \dots , N - 1 ,
\label{1_70}\\
	R_{N N} &=& - N \left(
		\beta^{\prime\prime} + {\beta^{\prime}}^2
	\right),
\label{1_80}
\end{eqnarray}
where the prime denotes differentiation with respect to $x^N$.
The structure of Eqs.~\eqref{1_20} coincides with the equations of Einstein's general relativity where the source of gravitational field is the effective energy-momentum tensor~\eqref{1_30}.

After substitution of the metric \eqref{1_50} and the components \eqref{1_60}-\eqref{1_80} in the modified equations~\eqref{1_20},
we have 
\begin{eqnarray}
	(1 - N) \beta^{\prime \prime } + \frac{N (1 - N)}{2} \beta^{\prime 2 } &=&
	- \left(
		\beta^{\prime \prime} + N {\beta^{\prime}}^2
	\right) f_R + 	\frac{1}{2} f
\nonumber \\
	&&
	+\left[
		4 N^2 \beta^{\prime}\beta^{\prime \prime \prime} +
		2 N (N^2 - 1) {\beta^{\prime}}^2 \beta^{\prime \prime} +
		2 N \beta + 2 N (N + 1) {\beta^{\prime\prime}}^2
	\right] f_{RR}
\nonumber \\
	&&
	+\left[
		4 N^2 {\beta^{\prime \prime \prime}}^2
		+ 8 N^2 (N + 1) \beta^{\prime} \beta^{\prime\prime}
		\beta^{\prime\prime\prime} +
		 4 N^2 (N + 1)^2 {\beta^{\prime}}^2 {\beta^{\prime \prime}}^2
	\right] f_{RRR}
\label{1_90} \\
	\frac{N (1 - N)}{2} \beta^{\prime 2} &=&
	- N \left(
		\beta^{\prime \prime} + {\beta^{\prime}}^2
		\right) f_R + \frac{1}{2} f +
	2 N^2 \beta^{\prime} \left[
	\beta^{\prime \prime \prime} + (N + 1) \beta^{\prime}	\beta^{\prime\prime}
	\right] f_{RR}.
\label{1_100}
\end{eqnarray}
Here, we use the designations $f_{R} = d f/d R$,
$f_{RR} = d^2 f/dR^2$, and $f_{RRR} = d^3 f/dR^3$.  Also, in the above equations,
we have taken into account that the scalar curvature is
$$   R = 2 N \beta^{\prime\prime} + N (N + 1) {\beta^{\prime}}^2.$$

Consider now Eq.~\eqref{1_100}, since, according to the Bianchi identity, Eq.~\eqref{1_90} follows from Eq.~\eqref{1_100}.
Dividing Eq.~\eqref{1_100} by the coefficients of the $\beta^{\prime \prime \prime}$,  we get the following equation:
\begin{equation}
\begin{split}
\label{1_120}
  & \beta^{\prime\prime\prime}
  - \dfrac{1}{n} \dfrac{{\beta^{\prime\prime}}^2}{\beta^{\prime}}
  + \dfrac{(N - 1)(1 + N - 2 n)}{4n (n - 1)} {\beta^{\prime}}^3
 + \dfrac{2 (N + 1) (n^2 + 1) - n (3 N + 5)}{2n(n-1)}
 \beta^{ \prime }\beta^{\prime \prime}
\\
 &
	 -\dfrac{N - 1}{4 \alpha N n (n-1)}
	\left[
		 2 N \beta^{\prime \prime}+N (N + 1) {\beta^{\prime}}^2
	\right]^{2-n} \beta^{\prime} = 0.
\end{split}
\end{equation}

Since this equation does not depend explicitly on the coordinate $x^N$, it is always
possible to shift the position of the brane to the point $x^N=0$ by the corresponding transformation of coordinates.
Then, in the neighborhood of this point, the solution of Eq.~\eqref{1_120} is sought in the form
\begin{equation}
	\beta\left(x^N\right) = \beta_0 +
	\gamma \left( x^N\right)^\delta + \dots ,
\label{1_130}
\end{equation}
where $\beta_0, \gamma $, and $\delta$ are constants.
In turn, without loss of generality, we may set $\beta_0 = 0$, corresponding to the redefinition of the coordinates
$e^{\beta_0} x^A \rightarrow x^A$.

The leading terms in Eq.~\eqref{1_120} are the terms
$\beta^{\prime \prime \prime}$ and $ \dfrac{1}{n}\dfrac{\beta^{\prime\prime2}}{\beta^{\prime}}$.
In order to {\it ensure} their finiteness, it is necessary to take $\delta > 3$.
Then, substituting the expansion \eqref{1_130} in \eqref{1_120}
and equating the coefficient of the
$
	\left( x^N \right)^{\delta - 3}
$ to zero, we get the following expression for the parameter $\delta $:
\begin{equation}
	\delta = \frac{2n - 1}{n - 1} .
\label{1_140}
\end{equation}
As a result, it is seen that the relationship between the parameters $\delta$ and $n$ does not involve the dimension of the bulk space.
Taking into account that $\delta > 3$, we get the following inequality for possible values of $n$:
\begin{equation}
	1 < n < 2 .
\label{1_150}
\end{equation}

\section{Numerical solutions}
\label{num_sol}

Apparently, it is impossible to find an analytic solution to Eq.~\eqref{1_120} over all of the space.
Numerical study of this equation for arbitrary dimension is impossible as well.
For this reason, we will carry out a numerical study for some specific dimensions.

For Eq.~\eqref{1_120}, there are four independent parameters which determine the solution: the dimension of spacetime $N$,
the parameters of modified gravity $n$ and $\alpha$,
and the parameter $\gamma$ which determines the magnitude of the metric function $\beta$ in the neighborhood of the brane.

In the simplest case one can consider solutions with even values of the parameter $\delta$,
which in turn is related to $n$ through the expression~\eqref{1_140}.
It is evident that in this case the solutions must be even functions with respect to $x^N$.

Much more complicated solutions do exist for an arbitrary value of the exponent $n$:
\begin{itemize}
\item Numerical analysis indicates that when $n = (2 p +1)/(2 q + 1)$ (here $p$ and $q$ are integer) and $\gamma>0$
a regular solution does exist for $x^N > 0$,
but for $x^N < 0$ it becomes singular. Our analysis shows that in the latter case there can exist regular solutions for $\gamma < 0$.
Then one can obtain a regular brane solution in all the space
by matching the regular solutions from the left and from the right of the brane
at the point $x^N = 0$. This can be done since for our choice of the exponent
$\delta$ the magnitude of the function $\beta$, as well as its first and second derivatives, are equal to zero on the brane:
$\beta(0) = \beta^\prime(0) = \beta^{\prime \prime}(0) = 0$ (the fixed point).
\item For irrational $\delta$, it is much more complicated to construct the solutions. The reason is that for $x^N < 0$,
near the origin of coordinates, the situation can occur when it will be necessary to calculate the degree of some negative number.
Subtracting a minus sign before such a number, the problem of calculating the number $(-1)^\delta$ for irrational $\delta$ occurs.
As is well known,
$
	(-1)^\delta = \exp\left( \imath m \pi \delta \right) =
	\cos \left( m \pi \delta \right) +
	\imath \sin \left( m \pi \delta \right)
$, where $m$ is an integer. In the general case, this is a complex number, and the solution apparently is absent.
\end{itemize}


\begin{figure}[t]
\begin{minipage}[t]{.3\linewidth}
	\begin{center}
		\includegraphics[width=1\linewidth]{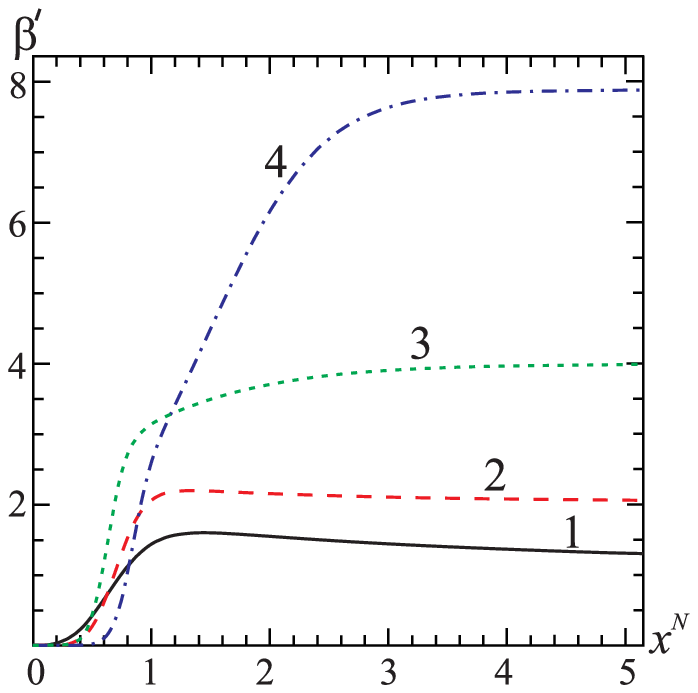}
	\end{center}
\vspace{-0.5cm}
\end{minipage}\hfill
\begin{minipage}[t]{.31\linewidth}
	\begin{center}
		\includegraphics[width=1\linewidth]{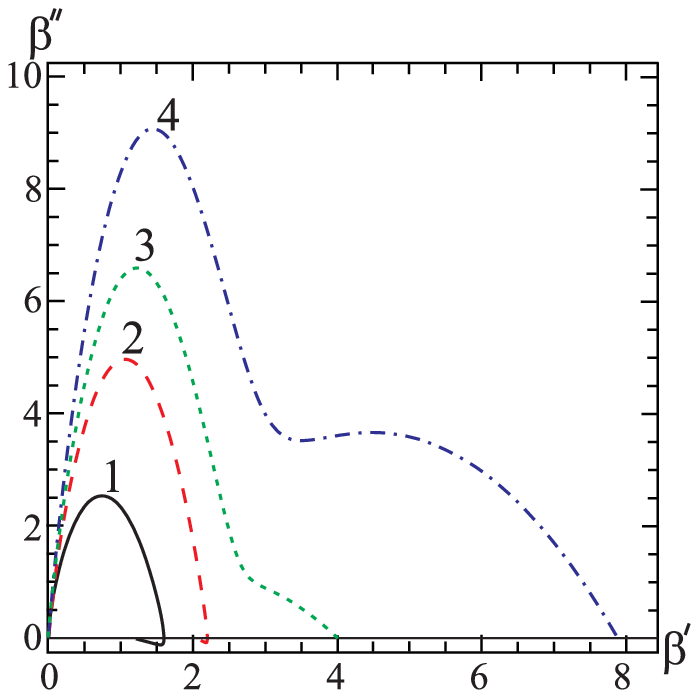}
	\end{center}
\vspace{-0.5cm}
\end{minipage}\hfill
\begin{minipage}[t]{.32\linewidth}
	\begin{center}
		\includegraphics[width=\linewidth]{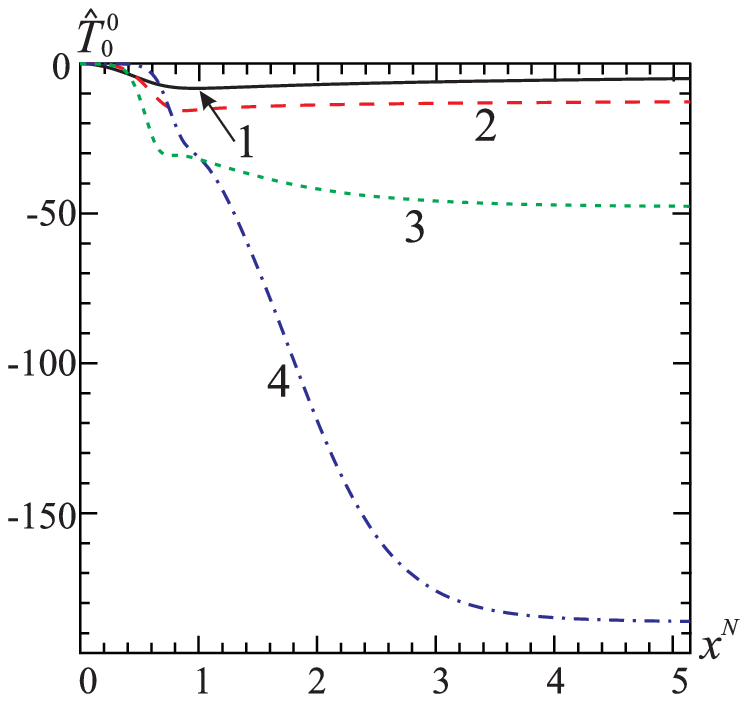}
	\end{center}
\vspace{-0.5cm}
\end{minipage}
\caption{The behavior of the function $\beta^\prime(x^N)$ (left panel), the phase portrait (middle panel), and the effective energy density $\hat T^0_0$  (right panel),
depending on different values of the parameter $\delta = 4, 6, 8, 10$ (or $n = 3/2, 5/4, 7/6, 9/8$)
for the curves 1, 2, 3, 4, respectively.
The graphs are plotted for the case of $N = 3$, $\alpha = 1$, $\gamma = 1$.
}
\label{delta}
\end{figure}

\begin{figure}[t]
\begin{minipage}[t]{.31\linewidth}
	\begin{center}
		\includegraphics[width=1\linewidth]{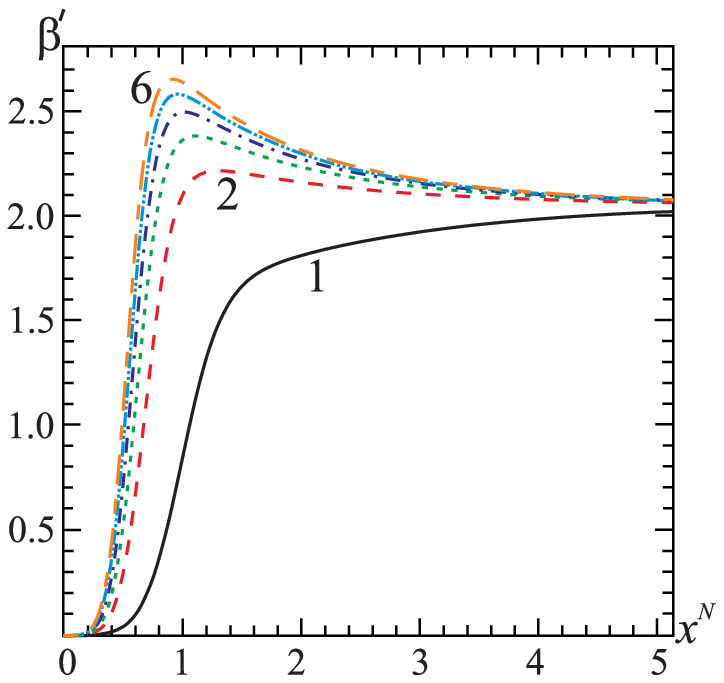}
	\end{center}
\vspace{-0.5cm}
\end{minipage}\hfill
\begin{minipage}[t]{.3\linewidth}
	\begin{center}
		\includegraphics[width=1\linewidth]{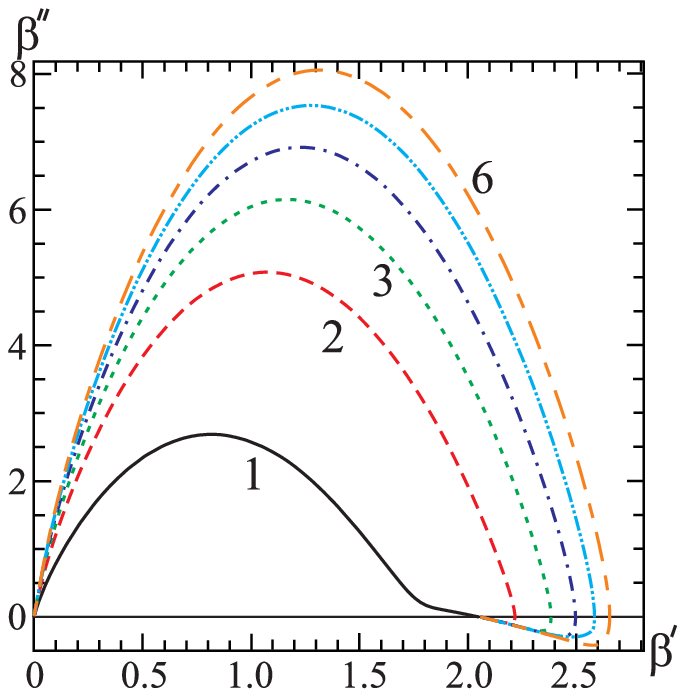}
	\end{center}
\vspace{-0.5cm}
\end{minipage}\hfill
\begin{minipage}[t]{.32\linewidth}
	\begin{center}
		\includegraphics[width=1\linewidth]{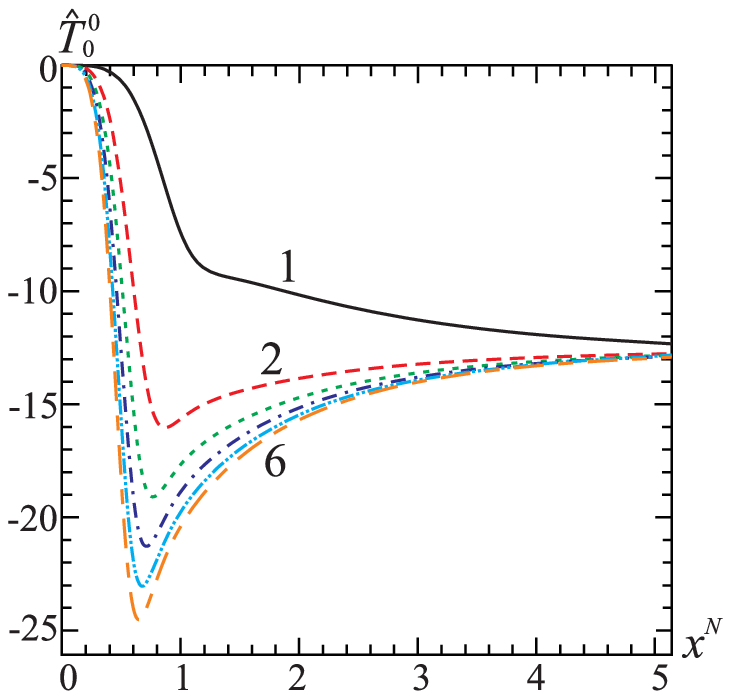}
	\end{center}
\vspace{-0.5cm}
\end{minipage}\hfill
\caption{The behavior of the function $\beta^\prime(x^N)$ (left panel), the phase portrait (middle panel), and the effective energy density $\hat T^0_0$ (right panel),
depending on different values of the parameter  $\gamma = 0.1, 1.0, 2.0, 3.0, 4.0, 5.0$
for the curves 1, 2, 3, 4, 5, 6, respectively (numbering is from the bottom to the top for the left and middle panels, and from the top to the bottom for the right panel).
For these plots,	$N = 3$, $\delta = 6$, $n = 5/4$, $\alpha = 1$.
}
\label{gamma}
\end{figure}

\begin{figure}[t]
\begin{minipage}[t]{.31\linewidth}
	\begin{center}
		\includegraphics[width=\linewidth]{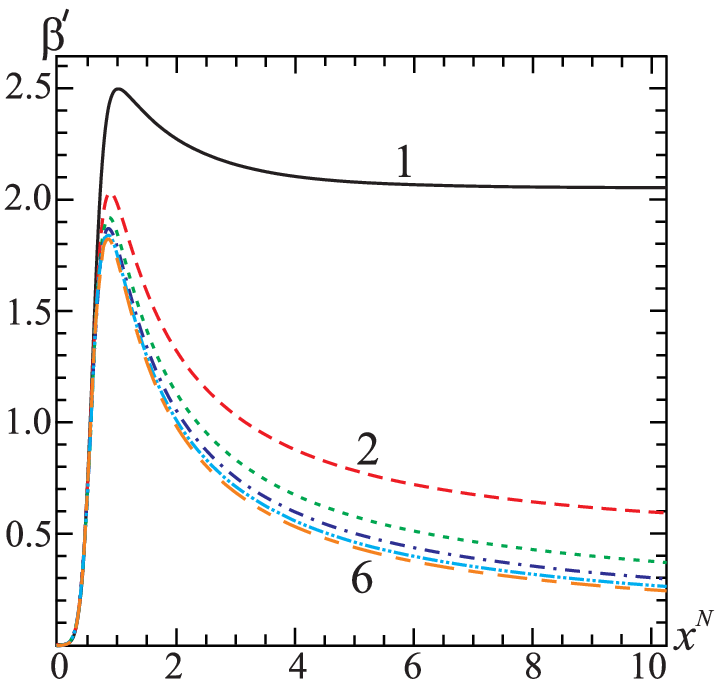}
	\end{center}
\vspace{-0.5cm}
\end{minipage}\hfill
\begin{minipage}[t]{.3\linewidth}
	\begin{center}
		\includegraphics[width=\linewidth]{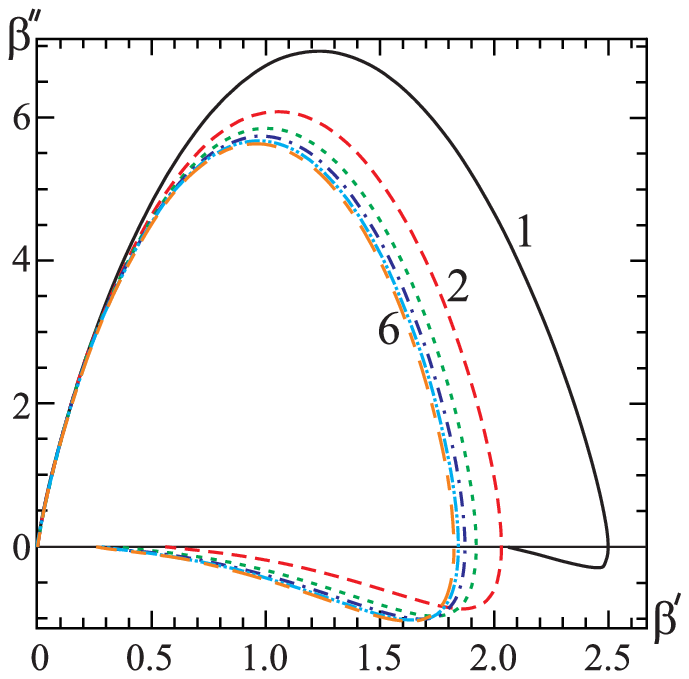}
	\end{center}
\vspace{-0.5cm}
\end{minipage}\hfill
\begin{minipage}[t]{.32\linewidth}
	\begin{center}
		\includegraphics[width=\linewidth]{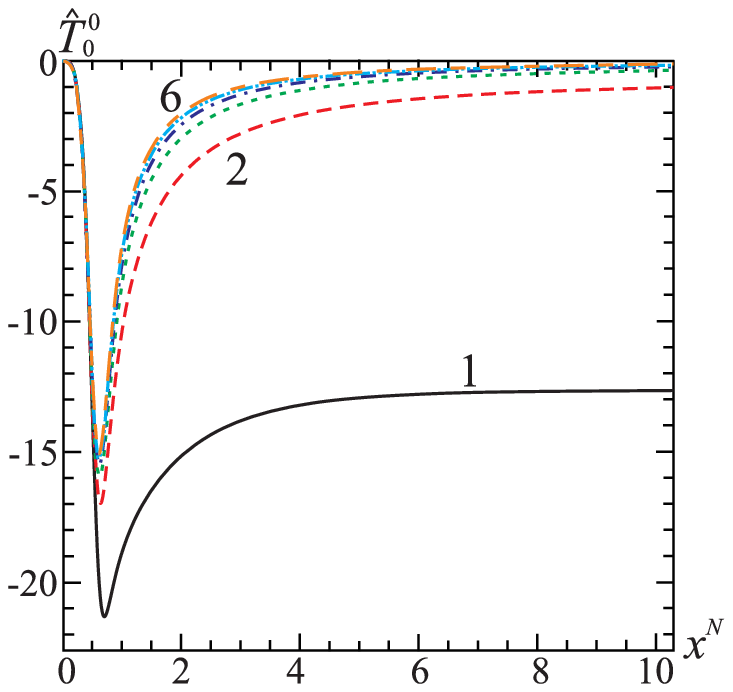}
	\end{center}
\vspace{-0.5cm}
\end{minipage}
\caption{The behavior of the function $\beta^\prime(x^N)$ (left panel), the phase portrait (middle panel), and the effective energy density $\hat T^0_0$ (right panel),
depending on different values of the parameter $\alpha = 1, 2, 3, 4, 5, 6$
for the curves 1, 2, 3, 4, 5, 6, respectively (numbering is from the top to the bottom for the left and middle panels, and from the bottom to the top for the right panel). For these plots,
	$N = 3$, $\delta = 6$, $n = 5/4$, $\gamma = 5$.
}
\label{alpha}
\end{figure}
\begin{figure}
\begin{minipage}[t]{.31\linewidth}
	\begin{center}
		\includegraphics[width=\linewidth]{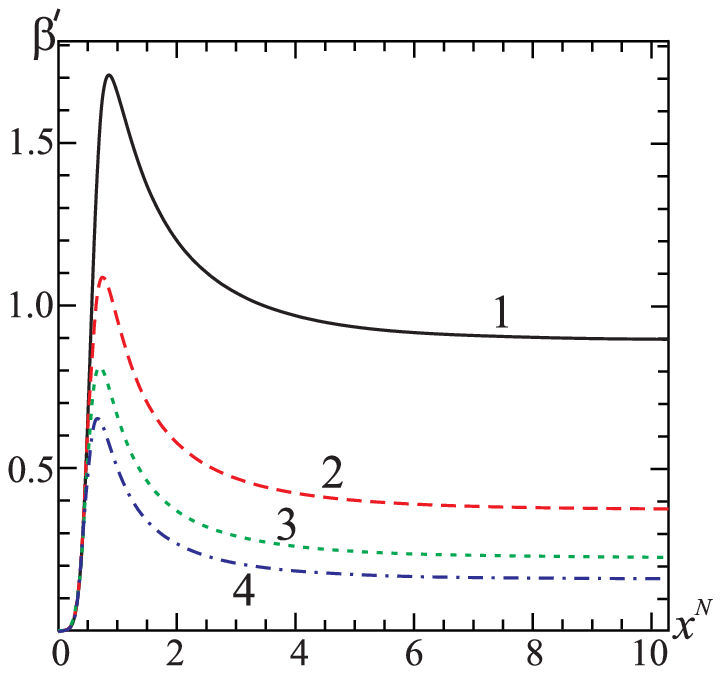}
	\end{center}
\vspace{-0.5cm}
\end{minipage}\hfill
\begin{minipage}[t]{.3\linewidth}
	\begin{center}
		\includegraphics[width=\linewidth]{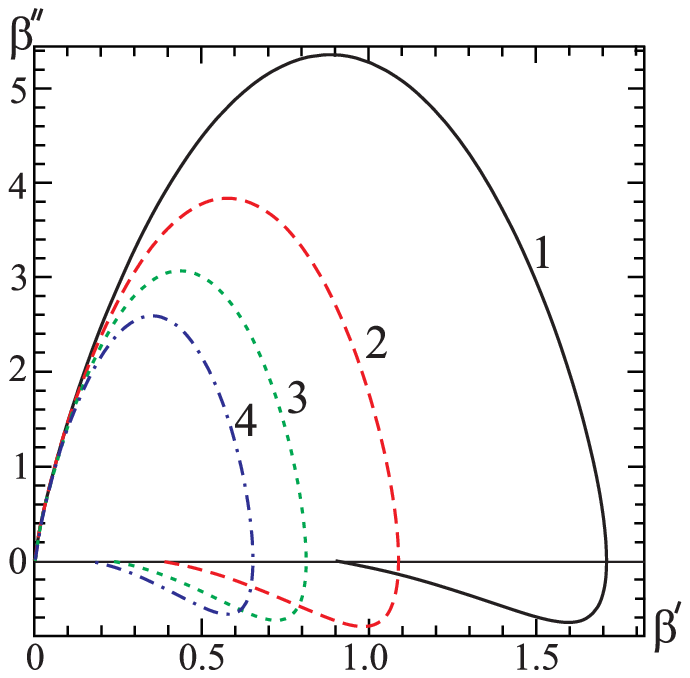}
	\end{center}
\vspace{-0.5cm}
\end{minipage}\hfill
\begin{minipage}[t]{.31\linewidth}
	\begin{center}
		\includegraphics[width=\linewidth]{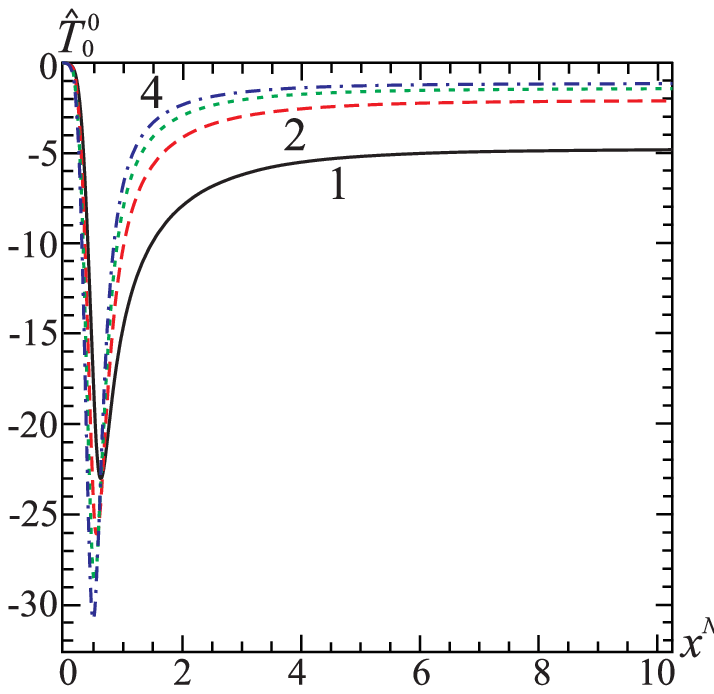}
	\end{center}
\vspace{-0.5cm}
\end{minipage}
\caption{The behavior of the function $\beta^\prime(x^N)$ (left panel), the phase portrait (middle panel), and the effective energy density $\hat T^0_0$ (right panel),
depending on different values of the spacetime dimension $N = 4, 6, 8, 10$
for the curves 1, 2, 3, 4, respectively. For these plots,
	$\delta = 6$, $n = 5/4$, $\gamma =3$, $\alpha = 1$.
}
\label{dim}
\end{figure}

Consistent with this, we consider here only the simplest case of regular, even plane symmetric solutions.
To demonstrate the properties of the systems under consideration,
we have plotted the graphs for the derivative of the metric function $\beta^\prime (x^N)$, the phase portrait
$\beta^{\prime \prime}(\beta^\prime)$, and the effective energy density  $\hat T^0_0$ from Eq.~\eqref{1_30}
for different values of the parameters $\delta$ (Fig.~\ref{delta}), $\gamma$ (Fig.~\ref{gamma}),
 $\alpha$ (Fig.~\ref{alpha}), and  $N$ (Fig.~\ref{dim}).
 [Notice that here we have taken into account the restriction imposed upon the parameter $n$ by Eq.~\eqref{1_150}.]
 Since all the solutions shown in these figures are even, we have plotted them only for $x^N>0$.

The above solutions have an anti-de Sitter asymptotic behavior of the form 
$$	\beta \approx \left\lbrace
		\frac{\left[ N (N + 1)\right]^{2 - n}}{ \alpha N (N - 2 n +1)}
	\right\rbrace^ {\frac{1}{2 (n - 1)}} \left| x^N \right|~,
$$
which is valid as $x^N\to \pm\infty$.
It is seen from this expression that there is the restriction $(N - 2 n +1)>0$ on the parameters $n$ and $N$ for which this
asymptotic solution is applicable.


Analysing the results shown in Fig.~\ref{alpha}, one can conclude that
when the parameter $\alpha$ increases, the solutions tend to saturation~-- all the curves tend to some limit.
This result can be easily explained: it is seen from Eq.~\eqref{1_120} that the last term goes to zero when $\alpha$ increases.
Without this term, this equation yields limiting solutions to which solutions of Eq.~\eqref{1_120} tend as $\alpha$ increases.

\section{Conclusion}
\label{conclusion}

We have obtained regular, even plane symmetric solutions within multidimensional $R^n$ modified gravity.
From the physical point of view, these solutions can describe thick branes with codimension~1.
The properties of such branes depend on four quantities: the parameter
$\gamma$ which describes the behavior of the solutions near the origin of coordinates, the parameters
$\alpha$ and $n$ related to the specific  type of modified gravity, and the dimension of spacetime~$N$.

In order to analyse the systems under consideration,
we have found the families of solutions for fixed values of the three above parameters and when the value of the fourth parameter has been changed.
As a result, we have shown that:
\begin{itemize}
\item Regular solutions may exist only for definite values of the parameter $n$ lying in the range $1<n<2$.
In this case all the derivatives of the metric function $\beta$ are equal to zero that permits one to
place the brane at the fixed point $x^N=0$ directly. The presence of the fixed point ensures the
existence of both $Z_2$-symmetric and nonsymmetric solutions.
\item Not all $n$ permit the existence of regular solutions:
	\begin{itemize}
	\item
	If $n = (2 p +1)/(2 q + 1)$ (where $p$ and $q$ are integer), the solutions are regular for $x^N > 0$ and may be singular for $x^N < 0$;
	\item
    If the exponent $n$ is an irrational number, in general, there are no solutions.
    \end{itemize}
\item According to Eq.~\eqref{1_20}, its right-hand side plays the role of the effective energy-momentum tensor $\hat T_{A}^B$.
We have studied the dependence of the effective energy density $\hat T^0_0$  on the values of the parameters  $\gamma, \alpha, n$, and $N$
and shown that it is always negative for all cases considered in the present paper.
This corresponds to the fact that all regular solutions obtained here describe an asymptotically anti-de Sitter spacetime.
\item When the parameter $\alpha$ increases, the solutions tend to some limiting solution that does not already depend on this parameter.
\end{itemize}

\section*{Acknowledgments}
V.D. and V.F. gratefully acknowledge support provided by Grant No.~BR05236322
in Fundamental Research in Natural Sciences by the Ministry of Education and Science of the Republic of Kazakhstan. They
are also grateful to the Research Group Linkage Programme of the Alexander von Humboldt Foundation for the support of this research
and would like to thank the Carl von Ossietzky University of Oldenburg for hospitality while this work was carried out.


\begin{thebibliography}{99}
\bibitem{Hashimoto:2012vsa}
  K.~Hashimoto, {\it D-brane: Superstrings and new perspective of our world} (Springer-Verlag, Berlin, Heidelberg, 2012).

\bibitem{Polchinski:1995mt}
  J.~Polchinski,
  Phys.\ Rev.\ Lett.\  {\bf 75}, 4724 (1995).

\bibitem{Horowitz:1991cd}
  G.~T.~Horowitz and A.~Strominger,
  Nucl.\ Phys.\ B {\bf 360}, 197 (1991).

\bibitem{arkani}
N. Arkani-Hamed, S. Dimopoulos, and G. Dvali,
Phys. Lett. {\bf B 429} 263 (1998);
I. Antoniadis, N. Arkani-Hamed, S. Dimopoulos, and G. Dvali,
Phys. Lett. {\bf B 436} 257 (1998).

\bibitem{gog}
M. Gogberashvili,
Int. J. Mod. Phys. {\bf D 11}, 1635 (2002); ibid. 1639 (2002);
M. Gogberashvili,
 Europhys. Lett., {\bf 49} (2000) 396;
M. Gogberashvili,
Mod. Phys. Lett., {\bf A 14} (1999) 202.

\bibitem{rs}
L. Randall and R. Sundrum,
Phys. Rev. Lett. {\bf 83}, 3370 (1999); ibid. 4690.

\bibitem{fermi1}
N. Arkani-Hamed and M. Schmaltz,
Phys. Rev. {\bf D 61}, 033005 (2000);
A. Mirabelli and M. Schmaltz,
Phys. Rev. {\bf D 61}, 113011 (2000).

\bibitem{fermi2}
S. Aguilar and D. Singleton, Phys. Rev. {\bf D 73}, 085007 (2006).

\bibitem{de}
C. Deffayet, G.R. Dvali, and G. Gabadadze,
Phys. Rev. {\bf D 65}, 044023 (2002);
M. Gogberashvili, Phys. Lett. {\bf B 636}, 147 (2006).

\bibitem{dm} M. Gogberashvili and M. Maziashvili,
Gen. Rel. Grav. {\bf 37},
1129 (2005).

\bibitem{Dzhunushaliev:2009va}
  V.~Dzhunushaliev, V.~Folomeev, and M.~Minamitsuji,
  Rept.\ Prog.\ Phys.\  {\bf 73}, 066901 (2010).

\bibitem{Schunck:2003kk}
  F.~E.~Schunck and E.~W.~Mielke,
  Class.\ Quant.\ Grav.\  {\bf 20}, R301 (2003)

\bibitem{Herdeiro:2017fhv}
  C.~A.~R.~Herdeiro, A.~M.~Pombo, and E.~Radu,
  Phys.\ Lett.\ B {\bf 773}, 654 (2017)

\bibitem{Dzhunushaliev:2019kiy}
  V.~Dzhunushaliev and V.~Folomeev,
  Phys.\ Rev.\ D {\bf 99}, no. 10, 104066 (2019)



\bibitem{Dzhunushaliev:2009dt}
  V.~Dzhunushaliev, V.~Folomeev, B.~Kleihaus, and J.~Kunz,
  JHEP {\bf 1004}, 130 (2010).

\bibitem{Zhong:2015pta}
  Y.~Zhong and Y.~X.~Liu,
  Eur.\ Phys.\ J.\ C {\bf 76}, no. 6, 321 (2016).

\bibitem{Dzhunushaliev:2019wvv}
  V.~Dzhunushaliev, V.~Folomeev, G.~Nurtaeva, and S.~D.~Odintsov,
  arXiv:1908.01312 [gr-qc].

\bibitem{DeFelice:2010aj}
  A.~De Felice and S.~Tsujikawa,
  Living Rev.\ Relativity\  {\bf 13}, 3 (2010).

\bibitem{Nojiri:2010wj}
  S.~Nojiri and S.~D.~Odintsov,
  Phys.\ Rep.\  {\bf 505}, 59 (2011).

\bibitem{Nojiri:2017ncd}
  S.~Nojiri, S.~D.~Odintsov, and V.~K.~Oikonomou,
  Phys.\ Rep.\  {\bf 692}, 1 (2017).




\end{thebibliography}
\end{document}